\documentclass[twocolumn]{aastex631}
\begin{document}

\title{Modeling multiband SEDs and light curves of BL Lacertae using a time-dependent shock-in-jet model}

\author[0000-0002-6941-7002]{Rukaiya Khatoon}
\affiliation{Centre for Space Research,
North-West University, 
Potchefstroom, 2520, South Africa}

\author[0000-0002-8434-5692]{Markus B\"ottcher}
\affiliation{Centre for Space Research,
North-West University,
Potchefstroom, 2520, South Africa}

\author[0000-0002-1173-7310]{Raj Prince}
\affiliation{Department of Physics, Institute of Science, Banaras Hindu University, Varanasi-221005, India}

\begin{abstract}
The origin of fast flux variability in blazars is a long-standing problem, with many theoretical models proposed to explain it. In this study, we focus on BL Lacertae to model its spectral energy distribution (SED) and broadband light curves using a diffusive shock acceleration process involving multiple mildly relativistic shocks, coupled with a time-dependent radiation transfer code. BL Lacertae was the target of a comprehensive multiwavelength monitoring campaign in early July 2021. We present a detailed investigation of the source's broadband spectral and light curve features using simultaneous observations at optical-UV frequencies with Swift-UVOT, in X-rays with Swift-XRT and AstroSat-SXT/LAXPC, and in gamma-rays with Fermi-LAT, covering the period from July to August 2021 (MJD 59400 to 59450). 
A fractional variability analysis shows that the source is most variable in gamma-rays, followed by X-rays, UV, and optical. This allowed us to determine the fastest variability time in gamma-rays to be on the order of a few hours. The AstroSat-SXT and LAXPC light curves indicate X-ray variability on the order of a few kiloseconds. Modeling simultaneously the SEDs of low and high flux states of the source and the multiband light curves provided insights into the particle acceleration mechanisms at play. This is the first instance of a physical model that accurately captures the multi-band temporal variability of BL Lacertae, including the hour-scale fluctuations observed during the flare.

\end{abstract}


\section{Introduction} \label{sec:intro}

Blazars are a specific class of radio-loud Active Galactic Nuclei (AGN), distinguished by the presence of a relativistic jet that is closely aligned with the observer's line of sight (\citealt{Urry_1995}). Their broad-band spectrum, ranging from radio to very high energy $\gamma$-rays, is dominated by the non-thermal emission from the powerful Doppler-boosted jets.
Blazars are further characterized by strong stochastic variability and occasional intense flares across the electromagnetic spectrum, ranging from microwave/radio to very high-energy gamma-rays, and occurring over a wide range of timescales spanning from minutes to years. Blazars are categorized into two main groups: flat spectrum radio quasars (FSRQs) and BL Lacertae objects. They are distinguished by the presence  (in FSRQs) or absence (in BL Lac objects) of broad optical emission lines with equivalent width $>$ 5 Angstrom (\citealt{Fan_2003}).

The broadband spectral energy distribution (SED) of blazars displays two distinct peaks. The lower energy peak typically extends from radio -- infrared (IR) to soft X-ray wavelengths, while the higher energy peak resides at MeV/GeV energies. The origin of the low-energy component is attributed to synchrotron emission, which arises from relativistic electrons spiraling around magnetic fields within the jet. Conversely, the high-energy component is often modeled as inverse-Compton (IC) scattering, where relativistic electrons interact with low-energy photons. These low-energy photons may originate either internally (through synchrotron emission) or externally (from sources like the accretion disk, broad-line region, dusty torus, etc.) to the jet. This modeling framework is commonly known as a leptonic model (\citealt{Ghisellini_2010}; \citealt{B13}; \citealt{Ghisellini_2014}). When the target photons for inverse-Compton scattering are synchrotron photons themselves, the process is termed synchrotron self-Compton (SSC; \citealt{MG85}; \citealt{Maraschi92};  \citealt{Sikora_2009}). Conversely, the scattering of external photons is referred to as the external-Compton (EC; \citealt{DSM92}; \citealt{DS93}; \citealt{Sikora94}; \citealt{Blazejowski2000}; \citealt{Ghisellini_1998}) process. Blazars are further categorized based on the location of the synchrotron peak ($\nu_s$), such as
low synchrotron
peaked blazar (LBL, $\nu_s$ $\leq$ 10$^{14}$ Hz), intermediate synchrotron
peaked blazar (IBL, 10$^{14}$ $\leq$ $\nu_s$ $\geq$ 10$^{15}$ Hz), and high synchrotron
peaked blazar (HBL, $\nu_s$ $\geq$ 10$^{15}$ Hz) \citep{Abdo10}.  

BL Lacertae (BL Lac) stands as the prototype of the BL Lac class of blazars, situated at a redshift of $z = 0.069$ (\citealt{Miller_1978}). While it is commonly categorized as LBL (\citealt{Nilsson_2018}), sometimes it is identified as an IBL (\citealt{Ackermann_2011}; \citealt{Hervet_2016}). 
BL Lacertae is renowned for its extensive variability across the electromagnetic spectrum, prompting numerous multi-wavelength campaigns by various groups (e.g., \citealt{Hagen_2002}; \citealt{Bottcher_2003}; \citealt{Marscher_2008}; \citealt{Raiteri2009}; \citealt{Raiteri2013}; \citealt{Wehrle2016}; \citealt{Magic_2019}; \citealt{Weaver_2020}) aimed at unraveling its properties in different
bands. Its intense variability spans diverse timescales, from minutes (\citealt{Ravasio_2002, Villata_2002, Meng_2017, Fang_2022}) to years (\citealt{Carini_1992, Villata_2009, Raiteri_2013}). Recently, BL Lacertae underwent a sequence of notable high-flux episodes spanning from 2020 to 2021 across the optical to very high-energy (VHE) $\gamma$-rays spectrum as reported in various Astronomical Telegrams (\citealt{ATEL_Grishina_2020, 2020ATel_cheung, 2020ATel_blanch, 2020ATel_ojha, 2021ATel_buson, 2021ATel_blanch, 2022ATel_lamura, 2022ATel_prince}), prompting several investigations into its multiwavelength characteristics.

The Large Area Telescope (LAT) on board the Fermi Gamma-ray Space Telescope (Fermi-LAT) has been conducting regular monitoring of BL Lacertae across a broad energy range of 0.1-300 GeV. Observations frequently indicate that during periods of flaring activity, the average $\gamma$-ray flux above 100 MeV can exceed 10$^{-6}$ ph cm$^{-2}$ s$^{-1}$ (for detailed references, see  \citealt{cutini_2012ATel, 2020ATel_cheung, mereu_2020ATel, 2020ATel_ojha, cutini_2021ATel, 2022ATel_lamura}). \citealt{pandey_2022} reported notable fluctuations in minute-scale binned $\gamma$-ray light curves, with a halving timescale of approximately 1 minute. On April 27, 2021, they recorded a peak 
$\gamma$-ray flux (0.1-500 GeV) of 2.0$\times$10$^{-5}$ ph cm$^{-2}$ s$^{-1}$, marking the historically brightest $\gamma$-ray flux ever detected from the source. This finding suggests that the $\gamma$-ray radiation is highly beamed and likely originates from an extremely compact region within the jet.

 In addition, BL Lacertae presents a fascinating subject for in-depth X-ray investigations since the X-ray spectrum resides at the convergence of the two broad components of the multiwavelength SED. X-ray observations conducted at various points in time reveal significant fluctuations in flux and spectral characteristics. These fluctuations suggest that the X-ray emission alternates between being dominated by the high-energy segment of synchrotron radiation and the low-frequency segment of the high-energy bump in the SED. 
 Notably, BL Lacertae has exhibited a concave profile on multiple occasions \citep[e.g.,][]{Ravasio_2002}. A recent study by \citealt{Prince_bllac_2021}, observing significant flaring activity of BL Lacertae across various wavelengths in 2020, reveals that the source underwent a spectral change in X-rays and at optical-UV frequencies. They observed that during the October 2020 flare, the synchrotron emission component peaks at approximately $\sim$ 10$^{16}$~Hz, characteristic of HSP type blazars. Conversely, during the low-flux state, the synchrotron emission peaks around $\sim$ 10$^{14}$~Hz, indicating an LSP/ISP type blazar. This suggests that the source has changed its spectral characteristics from ISP to HSP while transitioning from low to high flux states. 

One-zone blazar models generally represent rapid, relativistic particle acceleration as the instantaneous injection of a non-thermal particle population, typically a power-law with injection index $q$. Injection indices around $q \sim 2 - 2.5$ are often found consistent with particle acceleration by mildly relativistic shocks through the first-order Fermi acceleration mechanism, often also called diffusive shock acceleration, where particle gain energy by repeated shock crossings between the upstream and downstream plasmas \citep[e.g.,][]{Peacock81,KS87a,KS87b,Ellison90,ED2004,SB_2012}.  
 
 However, there are cases where the power-law index of the emitting electrons must be as hard as $\sim$ 1 to match the observed X-ray data. This has often been considered as support for magnetic reconnection as the dominant particle acceleration mechanism \citep[e.g.][]{Sironi2014,Sironi2015}. This occurs when the turbulent magnetic field behind the shock front annihilate and convert magnetic energy to particle energy, leading to systematic fast variability and hard lags. \citet{Guo2014} demonstrated through simulations that in highly magnetized environments ($\sigma >> $ 1), the particle spectral index approaches to 1.0, necessary for longer particle injection times compared to first-order Fermi acceleration. 
 The radition and polarization signatures of magnetic reconnection in blazar jets have been extensively studied by \cite{Zhang2018,Zhang2020,Zhang2022}.
 
 Alternatively, kink instabilities in the jet leading to relativistic  particle acceleration. \cite{Zhang2017} studied the polarization signatures of this process, finding that it might lead to multi-wavelength flares and polarization-angle  swings.  \cite{Jorstad2022} invoked this process to interpret an  outburst of BL Lac in 2020 at 43 GHz, inferring a tight helical magnetic field resulting in current-driven kink QPOs, as previously predicted to occur by \cite{Dong2020}. 

 Finally, relativistic shear boundary layers have also been considered as a plausible site of relativistic particle acceleration \citep[e.g.,][]{Liang2013,Alves2014,Liang2017}. In particular, \cite{Liang2018,CB2024} showed that this acceleration mechanism leads to significantly more strongly beamed particle and radiation patterns compared to the standard relativistic beaming of an intrinsically isotropic radiation field in the co-moving frame of a relativistically moving radiation zone.
 This might offer an elegant solution to the long-standing Doppler factor crisis \citep{LL2010}. 
 
 While simultaneous broadband spectra offer valuable insights into constraining blazar jet models, significant uncertainties persist in interpreting them regarding the primary electron cooling, injection, and acceleration mechanisms. This challenge was starkly exemplified in the case of W Comae by \citet{BMR2002}. It is demonstrated that combining broadband spectra with timing and spectral variability data, alongside time-dependent model simulations \citep{bottcher_chiang_2002, Krawczynski_2002}, can help resolve some of these ambiguities. 

In early July 2021, BL Lacertae underwent a multiwavelength outburst as reported in various Astronomer's Telegrams (ATel\# 14751, \citealt{2021ATel14751....1H}; ATel\# 14773, \citealt{2021ATel14773....1C}; ATel\# 14774, \citealt{2021ATel14774....1P}; ATel\# 14777, \citealt{2021ATel14777....1B}; ATel\#14782, \citealt{2021ATel14782....1P} ATel\# 14783, \citealt{2021ATel14783....1C}; ATel\# 14820, \citealt{2021ATel14820....1K}; ATel\# 14826, \citealt{2021ATel14826....1B}; ATel\# 14839, \citealt{2021ATel14839....1P}). Subsequently, we proposed follow-up target of opportunity observations in the optical/UV and X-ray bands using telescopes such as Swift-XRT/UVOT, AstroSat, and NuSTAR. In this study, we perform a comprehensive investigation of the multiband spectral and light curve characteristics of BL Lacertae utilizing data obtained from these telescopes, in conjunction with gamma-ray data from Fermi-LAT, spanning from July 2021 (MJD 59400) to August 2021 (MJD 59450).

The structure of the paper is as follows: In \S 2, we outline the multi-wavelength observations and the procedure for data analysis. In \S 3, we present the broadband light curve and identify different flux states, followed by the findings from the temporal and 
spectral analysis of the observations. \S 4 delves into the comprehensive broadband spectral as well as lightcurve modeling of the source.
 Finally, \S 5 discusses and summarizes the findings of the study.

\section{Data reduction and analysis}
The long-term flaring period in 2021 was continuously monitored by the Fermi-LAT in $\gamma$-rays along with X-ray and optical-UV by Swift-XRT and UVOT.
We also obtained pointing observations of the flaring and low states by the  AstroSat and NuSTAR telescopes. A very-high-energy $\gamma$-ray flare was also observed by the CTA LST-1 and MAGIC telescopes, but we do not have access to those data. 

\subsection{Fermi-LAT}
The Large Area Telescope (LAT) is the main gamma-ray detector on the Fermi satellite, launched by NASA in 2008 \citep{Atwood_2009}. It operates within the energy range of 20 MeV to over 300 GeV and scans a wide field of view of approximately 2.4 steradians. It takes around 3 hours to conduct a full scan of the sky. BL Lacertae is a source regularly monitored by Fermi-LAT.

To investigate the gamma-ray behavior of BL Lacertae, we conducted an analysis using data collected from July 5 to August 24, 2021 (MJD 59400-59450) covering multiple flares. This period encompasses both low and high flux states of the source. Our analysis focused on a 10-degree region of interest (ROI) centered on the source position, following the standard analysis procedures outlined in the Fermi science tools (version v10r0p5).

We performed the analysis within the energy range of 100 MeV to 300 GeV, employing evclass=128 and evtype=3 parameters, and applying a zenith angle cut of $>$90$^{\circ}$ to minimize contamination from gamma-rays originating from the Earth's limb. The model XML file was generated using the ``\emph{$gll_-iem_-v07$}" galactic diffuse emission model and the ``\emph{$iso_-P8R3_-SOURCE_-V3_-v1$}" isotropic background model, both publicly available at the Fermi Science Support Center (FSSC).

Spectral models and parameters for sources within the ROI were taken from the fourth Fermi source catalog (4FGL; Abdollahi et al., 2020). We employed the maximum likelihood method to optimize the spectral parameters of these sources, with detection significance quantified using the test statistics (TS). Sources with TS $<$ 9 \citep[-approximately 3$\sigma$ significance;][]{Mattox_1996} were excluded from further analysis. This process was conducted using the ``unbinned likelihood analysis with Python" developed by the Fermi collaboration.

The default spectral model for BL Lacertae was a log parabola, with model parameters optimized through likelihood analysis. Subsequently, we fitted the source spectrum integrated over the entire energy range of 0.1-300 GeV during the period MJD 59400 -- 59450 using a log-parabola function. During light curve generation, the parameters of all sources except BL Lacertae were frozen, resulting in 6-hour binned gamma-ray light curves for flux and spectral parameters.

\subsection{Neil Gehrels Swift Observatory}

The Neil Gehrels Swift Observatory, launched in 2004, is a space-based facility designed to detect and study bright transients, such as gamma-ray bursts, across various wavelengths, including optical-UV and X-rays. This multiwavelength mission is equipped with three specialized instruments: an ultra-violet optical telescope (UVOT), an X-ray Telescope (XRT), and a burst alert telescope (BAT), enabling it to focus on optical-UV, soft X-ray, and hard X-ray energies. 

The source BL Lacertae has been regularly observed through both routine monitoring and targeted observations under the target of opportunity (ToO) program, using XRT and UVOT. Numerous observations were carried out during the flaring phase to monitor the object's X-ray and optical-UV activity. For this particular study, we utilized Swift observations totaling an exposure time of 411.2~ks, covering a period similar to that considered for Fermi-LAT (MJD 59400 to 59450). Below, we present the observations and detailed analysis.

\subsubsection{XRT} 

The X-ray Telescope (XRT) operates within the energy range of 0.3-10 keV. To reduce the X-ray data, we followed the standard procedure. Initially, raw data was obtained from the HEASARC webpage, and cleaned event files were generated employing the standard XRTPIPELINE. In this analysis, we used recent calibration files such as CALDB with version 20200305. The cleaned event files for the photon counting (PC) mode were subsequently employed to generate both source and background spectra, using the XSELECT tool.  

To extract the source spectrum, a 20 arcsec circular region is chosen around the source, while for the background spectrum, a 40 arcsec circular region is selected far from the source to avoid contamination. In instances of pile-up-affected observations where the source count rate exceeds 0.5 ct/sec, an annular region with inner and outer radii of 4 arcsec and 20 arcsec, respectively, is considered as the source region. Ancillary response files (ARF) and redistribution matrix files (RMF) are generated using XRTMKARF from the Swift CALDB.  
The grppha tool is used to merge the source spectrum, background spectrum, as well as ARF and RMF files, resulting in a grouped spectrum used for modeling. The X-ray spectrum corresponding to each observation is fitted with an absorbed power-law model using XSPEC \citep{Arnaud_1996}, from which flux and index estimates are obtained. A Swift-XRT light curve is then generated, with each point representing an observation. The X-ray spectra from high and low states are employed for broadband SED modeling.

\subsubsection{UVOT}

UVOT \citep{Roming_2005} operates across the optical and ultraviolet segments of the spectrum with its set of three optical filters (U, B, V) and three UV filters (W1, M2, W2). The instrument provides image files directly. To derive magnitudes from these images, the UVOTSOURCE task is employed, utilizing source and background regions of 5 and 10 arcseconds, respectively. The observed magnitudes are subsequently adjusted for Galactic extinction, employing E(B - V) = 0.0174 mag and the $A_V$/E(B - V ) ratio of 3.1 from \cite{SF11}. Conversion of magnitudes to flux units is accomplished using photometric zero-points sourced from \cite{Breeveld11} and conversion factors from \citealt{Larionov_2016}.

\subsection{AstroSat}
AstroSat is a multiwavelength space observatory equipped with five scientific instruments, encompassing a broad spectrum of energies from ultraviolet (UV) to hard X-rays. The instruments utilized in this study comprise the Soft X-ray Focusing Telescope (SXT) and the Large Area X-ray Proportional Counters (LAXPC).  AstroSat conducted observations of BL Lacertae under Target of Opportunity (ToO) proposals, with both Level-1 and Level-2 data accessible to the public through the ISRO Science Data Archive. In this study, we analyzed orbit-wise Level-1 data for the period of MJD 59406-59407 from the SXT and LAXPC instruments.

\subsubsection{AstroSat-SXT}
The SXT is a focusing telescope that enables X-ray imaging and spectroscopy sensitive mainly in 0.3 -- 7.1 keV energy band (\citealt{singh_2017}) 
We first processed the Level-1 SXT data for BL Lacertae collected for every individual orbit using the sxtpipeline task within the SXT software package AS1SXTLevel2, version 1.4b, and subsequently merged them into a cleaned event file utilizing the SXTEVTMERGER tool. Access to the analysis software and related tools can be found on the SXT Point of Contact (POC) website. 

The processed Level-2 cleaned event files were used to extract the source lightcurves, images, and spectra employing the XSELECT package (version 2.4d) integrated into HEAsoft. The cleaning process involved eliminating any contamination from charged particles resulting from the satellite's passage through the South Atlantic Anomaly region and occultation by the Earth, as well as selecting events with grade 0 -12, i.e., single-quadruple events. A ``pha cutoff" filter was applied to select energy channels (e.g., channel 80-700 for the 0.8-7.0 keV band) for the light curves. 

For spectral product generation, a circular region with a radius of 16 arcminutes, encompassing over 95~\% of the source pixels, was considered centered on the source position. To avoid issues arising from the large Point Spread Function of SXT, a standard background spectrum (``$SkyBkg_-comb_-EL3p5_-Cl_-Rd16p0_-v01.pha$") extracted from a composite product derived from a deep blank sky observation was utilized. Ancillary response files (ARF) for individual sources were generated using the sxtmkarf tool. Furthermore, a standard response file (``$sxt_-pc_-mat_-g0to12.rmf$") was obtained from the SXT POC website and employed as the RMF. Subsequently, the extracted source spectra were binned using the grppha tool to ensure a minimum of 60 counts per bin.

\subsubsection{AstroSat-LAXPC}
The LAXPC instrument is comprised of three proportional counter units, namely LAXPC10, LAXPC20, and LAXPC30, which cover the hard X-ray band from 3 to 80 keV \citep[][]{2016SPIE.9905E..1DY, 2017ApJS..231...10A}.
Among these detectors, LAXPC30 was deactivated due to a gain instability issue caused by gas leakage, while LAXPC10 operated at a low gain during the observation. Consequently, our analysis solely utilizes data from LAXPC20. Processing of the Level-1 data was carried out using the LaxpcSoft software package. Event files and standard GTI (Good Time Interval) files were generated using the built-in modules $laxpc_-make_-event$ and $laxpc_-make_-stdgti$, respectively, within the LaxpcSoft package. For faint sources such as AGNs, the background signal can dominate over the source counts.  Therefore, background estimation was performed within the 50 to 80 keV energy range, where the background appears relatively stable. Subsequently, appropriate background models and response functions were applied to generate spectra and light curves for the source employing the tools $laxpc_-make_-spectra$ and $laxpc_-make_-lightcurve$.

\subsection{NuSTAR}
Following the bright flaring activity in $\gamma$-rays and X-rays, we proposed a target of opportunity observation in hard X-rays with NuSTAR. The roughly 20 ksec observation was performed on 2021-08-16, starting at 02:06:09 (Obs Id: 90701628002), when the source was already in the low flux state. The data reduction is done using the latest NuSTAR data analysis software (NuSTARDAS) version 1.9.2 provided by
HEASOFT. To extract the source spectrum and the background
spectrum, a circular region of 20'' and 50'' was chosen around
the source and away from the source, respectively. We used the tool \texttt{nuproduct} to create the source and background spectrum files for both instruments (FPMA \& FPMB) and their corresponding RMF and ARF files. Both spectra were modeled in Xspec and the spectrum data points are used in the broadband SED modeling. 

\begin{figure*}
\centering
\includegraphics[ scale=0.6]{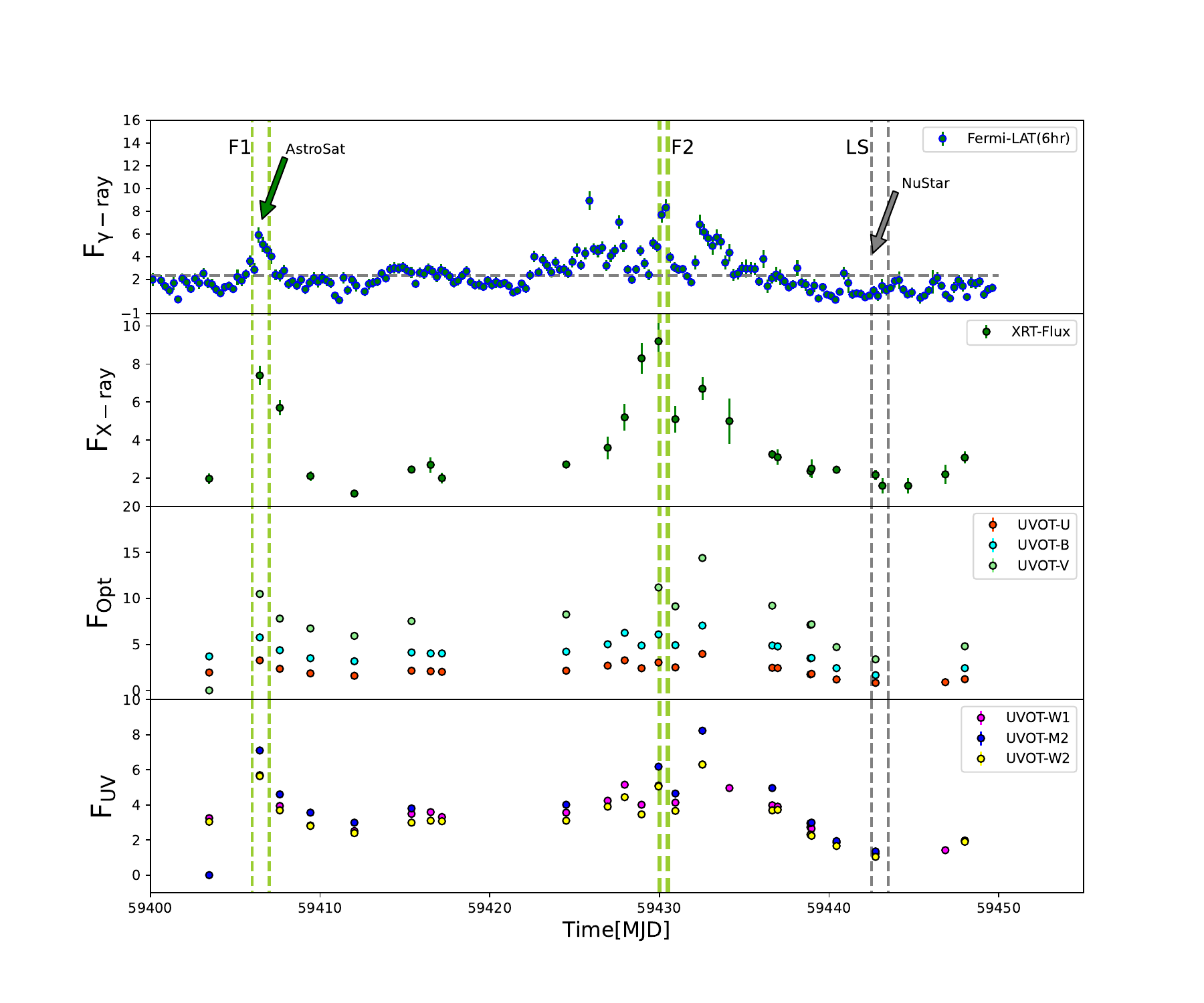}
\caption{Multiwaveband flux lightcurves of BL Lac between MJD 59400 and 59450. Panel 1: Fermi-LAT flux in 10$^{-6}$ ph cm$^{-2}$ s$^{-1}$, panel 2: Swift-XRT flux in 10$^{-10}$ erg cm$^{-2}$ s$^{-1}$, panel 3 \& 4: Swift-Optical/UV flux in 10$^{-10}$ erg cm$^{-2}$ s$^{-1}$.} 
\label{fig:mwl_lc}
\end{figure*}

 \begin{figure}
       \includegraphics[scale=0.45]{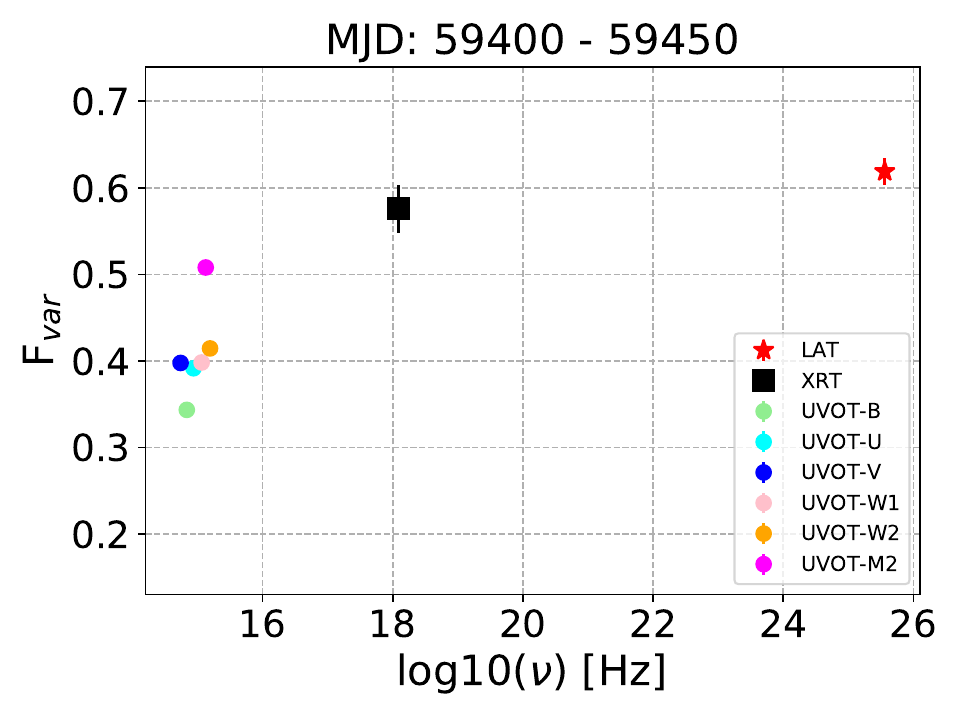}\label{freq-fvar}
       	\caption{Fractional variability vs. frequency for each light curve (see Figure \ref{fig:mwl_lc}). The numerical values are reported in Table \ref{tab:fvar}.}
	   \label{fig:freq-Fvar}
\end{figure}

\section{Multiwavelength lightcurves}
Figure \ref{fig:mwl_lc} shows the multiband light curves of BL Lac over $\sim$50 days (MJD 59400-59450), indicating variations in flux levels across all wavebands, including both high and low flux states. The panels in Figure \ref{fig:mwl_lc}, arranged from top to bottom, represent gamma-ray, X-ray, optical, and UV observations, respectively. In the top panel, the average flux (F$_{a}$ = 2.35$\times10^{-6} phcm^{-2}s^{-1}$) derived from 6-hour binned gamma-ray data is represented by a grey dashed horizontal line. We define the active flux state of the source as periods when the flux exceeds F$_{a}$. Consequently, we designate the periods MJD 59406-59407 and MJD 59429.5-59430.5 as the active states, during which the source exhibited two major outbursts in both gamma-ray and X-ray bands. These periods are referred to as flaring-1 `F1' and flaring-2 `F2', and represented by green dashed vertical lines, respectively. The AstroSat observations of $\sim$ 20 ks exposure were aligned with the outburst during F1 (MJD 59406-59407), facilitating variability analysis.  The lightcurve, when compared with F$_{a}$, indicates that the source remained in a low flux state for an extended duration; however, for our study we specifically selected the low-flux period of MJD 59442.5-59443.5, labeled as `LS', demarcated by grey dashed vertical lines. We chose this `LS’ period because we have NuSTAR observations alongside Fermi-LAT and Swift, which will help constrain the SED during the low-flux state.
Notably, during the F1 and F2 states, the source exhibits elevated flux levels in the X-ray, optical, and UV bands, whereas during the `LS' period, the source displays low flux levels across all wavebands.
 
\subsection{Variability study:}
We assess the degree of variability in the source by employing the fractional RMS variability amplitude, which is determined as the square root of excess variance ($\sigma_{XS}$). It considers the variability influenced by measurement uncertainties, and its mathematical representation is provided as follows \citep{Vaughan_2003}:

\begin{equation}\label{eq:fvar}
 F_{\rm var} = \sqrt{\frac{S^2 - {\sigma_{err}}^2}{{x_{av}}^2}},
\end{equation}

Here, $x_{av}$ represents the average flux, ${\it \sigma_{err}}^2$ denotes the mean square error in the observed flux, and {\it  S$^2$} signifies the total variance of the observed light curve. The error associated with {\it F$_{var}$} can be approximated as \citep{Vaughan_2003}:

\begin{equation}
err(F_{var}) = \sqrt{ \Big(\sqrt{\frac{1}{2N}}. \frac{\sigma_{err}^2}{{x_{av}}^2F_{var}} \Big)^2 + \Big( \sqrt{\frac{{\sigma_{err}}^2}{N}}. \frac{1}{x_{av}} \Big)^2 } 	
\end{equation}

Here, {\it N} represents the number of data points in the light curve. The fractional variability amplitude ({\it F$_{var}$}) calculated during the period spanning from MJD 59400 to MJD 59450, is reported in Table \ref{tab:fvar}, and is plotted {\it F$_{var}$} vs. photon frequency in Figure \ref{fig:freq-Fvar}. It is observed that BL Lac exhibits the highest variability in the gamma-ray and X-ray bands, followed by UV and optical wavelengths.

The source variability can also be characterized by the flux doubling/halving timescale, denoting the duration within which the flux changes by a factor of two between consecutive time intervals, expressed as \citep{Zhang_1999}
\begin{equation}
\centering
 t_{d} = \bigg|\frac{( f_{1} + f_{2})(t_2 - t_1)}{2( f_{2} - f_{1})}\bigg|
\end{equation}
Here, {\it t$_1$} and {\it t$_2$}  represent two consecutive times with fluxes {\it f$_{1}$}, and  {\it f$_{2}$}, respectively. The shortest variability time (t$_{\rm var}$) corresponds to the minimum value of $t_d$ estimated from all consecutive time intervals across the entire light curve. 

We observed the fastest $\gamma$-ray variability time (t$_{var}$) during the F1 state to be approximately 0.237$\pm$0.02 days (5.68$\pm$0.48 hours). The corresponding flux values were 2.84$\pm$0.59 and 5.90$\pm$0.69 ($\times10^{-6}\,ph\,cm^{-2}\,s^{-1}$) on MJD 59406.125 and MJD 59406.375, respectively. For the F2 state, the t$_{var}$ was found to be 0.233$\pm$0.01 days (5.60$\pm$0.24 hours) with flux values of 8.32$\pm$0.72 and 3.95$\pm$0.46 ($\times10^{-6}\,ph\,cm^{-2}\,s^{-1}$) on MJD 59430.375 and MJD 59430.625, respectively.

We derived the size of the emission region as R $\lesssim$ c t$_{\rm var} \delta/(1+z)$, where $\delta$ represents the jet Doppler factor and c denotes the speed of light, while t$_{\rm var}$ signifies the fastest variability timescale. With a t$_{\rm var}$ of $\sim$ 5.6 hours and $\delta \sim$ 15 (\citealt{BB_2019}), R is estimated to be $\sim 10^{16}$ cm. Similarly, the distance of the emission region can be estimated as d $\sim$ 2 c t$_{var}$ $\delta^2$/(1+z) $\sim$ 3.0$\times 10^{17}$ cm (\citealt{Abdo_2011}). These results are discussed in the SED modeling section, \S 4. 

\subsection{Correlation study:}

\begin{figure*}
    \centering
    \includegraphics[scale=0.36]{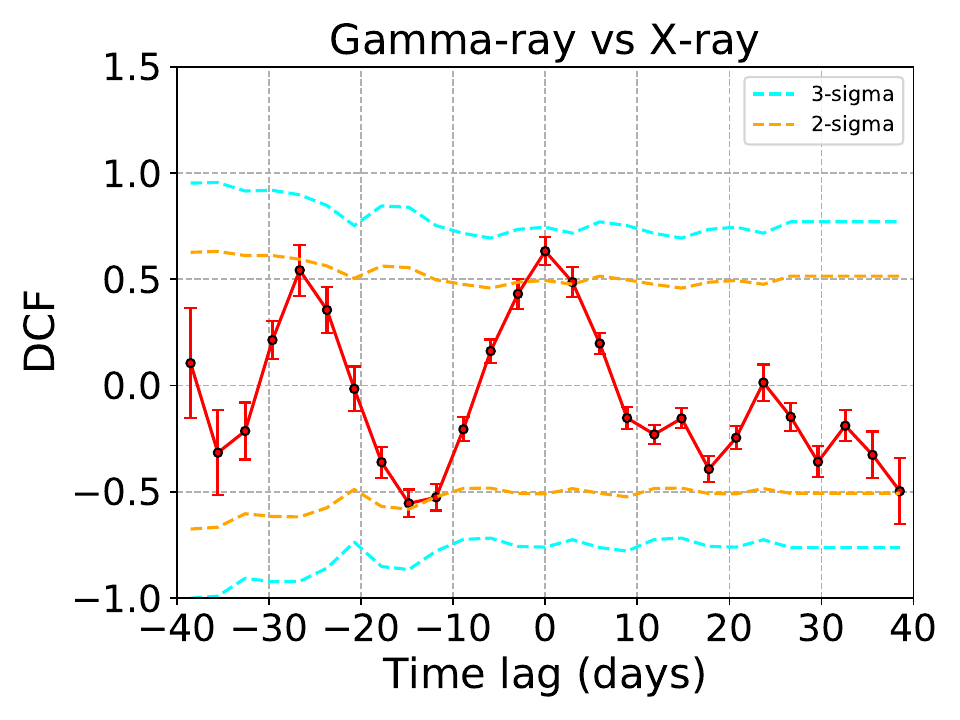}
    \includegraphics[scale=0.36]{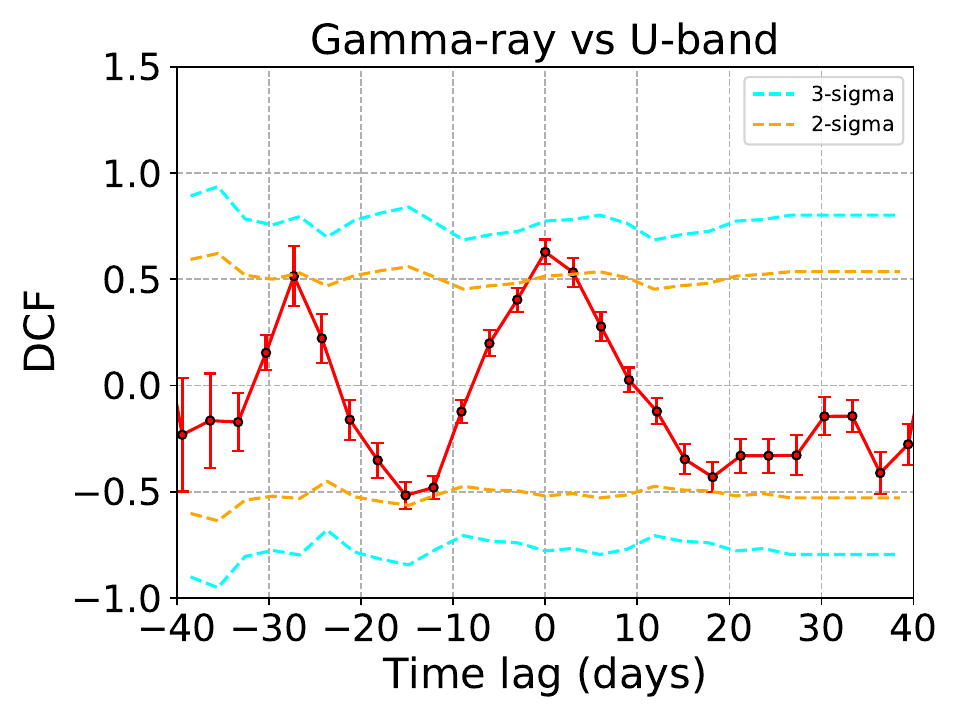}
    \includegraphics[scale=0.36]{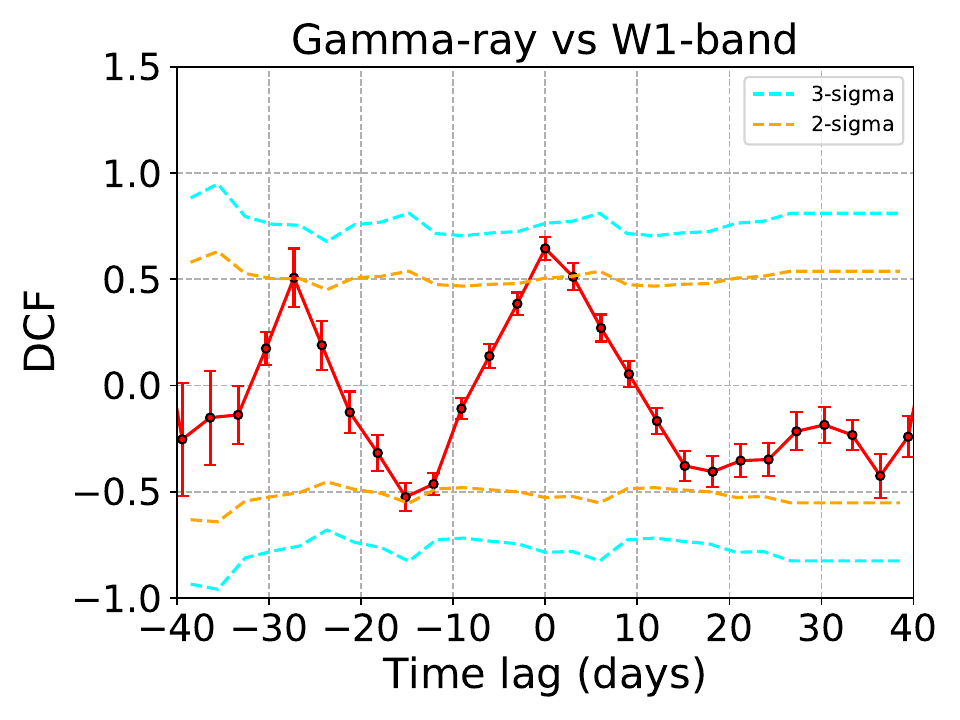}
    \includegraphics[scale=0.36]{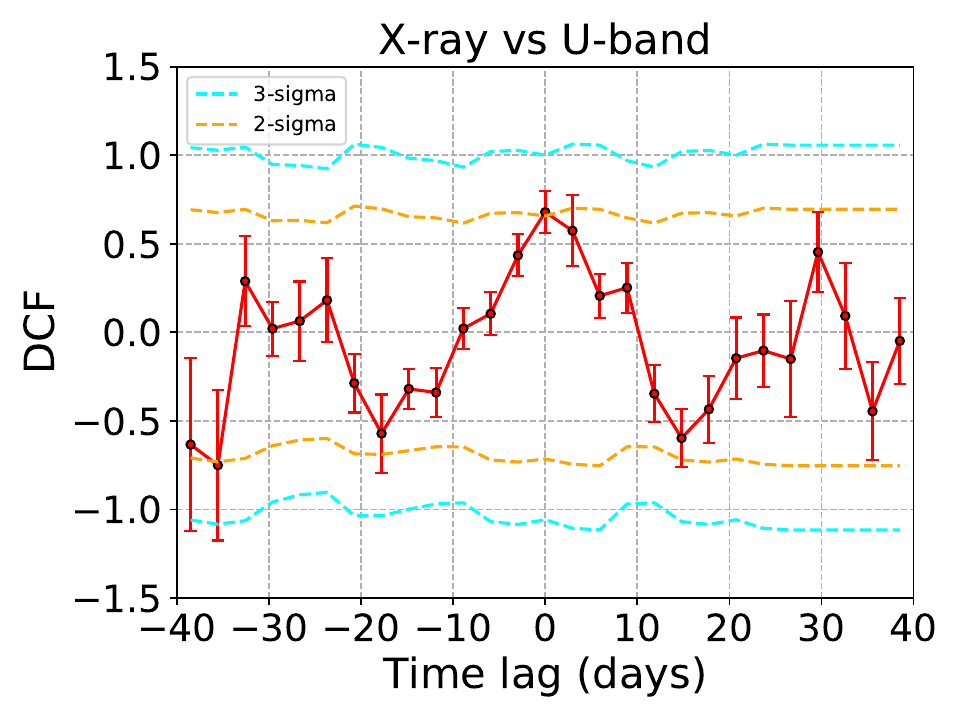}
    \includegraphics[scale=0.36]{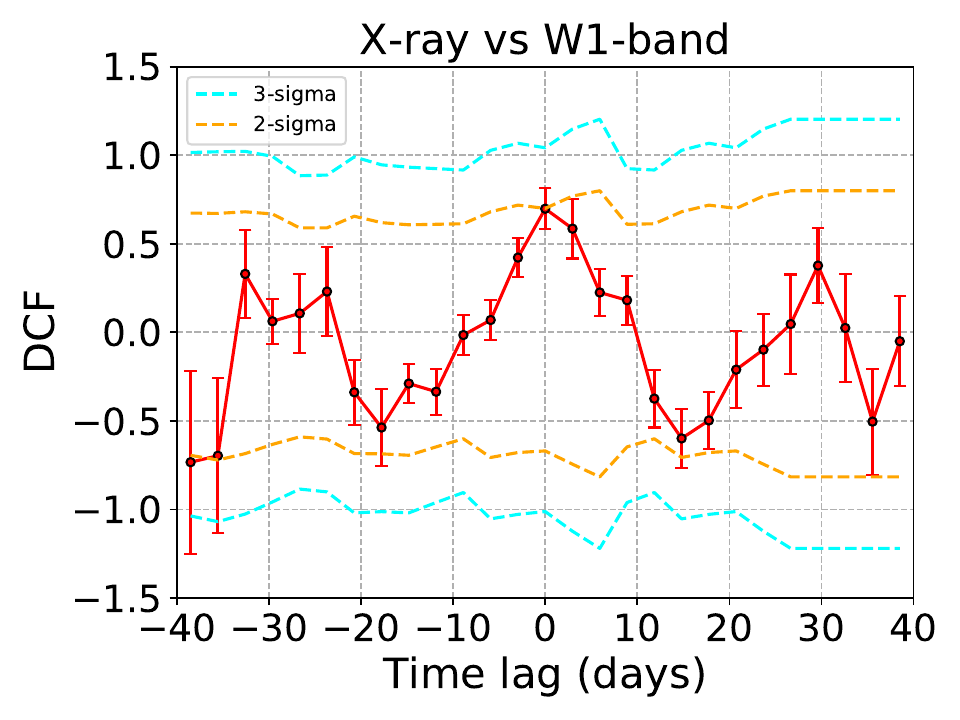}
    \caption{Correlations between various wavebands. Cyan and orange horizontal lines are 1 and 2-sigma significance contours. In all the cases zero time lags have been retrieved.}
    \label{fig:dcf}
\end{figure*}

To investigate the correlation between simultaneous $\gamma$-ray, X-ray, optical, and UV emission (Figure \ref{fig:mwl_lc}), we employed the Discrete Correlation Function (DCF) technique introduced by \cite{EK88}. The DCF analysis as shown in Figure \ref{fig:dcf} indicates a moderate correlation of approximately 58\% with zero time lag across all emissions in different wavebands. This zero time lag correlation suggests that the emissions are co-spatial and likely originate from the same particle population. The co-spatial nature of the broadband emission motivates us to use a single emitting zone to model the broadband SED and the associated light curves (see \S 4).

We also assessed the significance of the DCF peaks observed in the cross-correlation. To do this, we simulated 1000 light curves for each band following the method described by \cite{Emmanoulopoulos_2013} and implemented in a code by \cite{Connolly_2016}. The simulated light curves for each band were cross-correlated with those from other bands. The significance levels of 2$\sigma$ and 3$\sigma$ for each time lag were calculated and are displayed in orange and cyan, respectively, in Figure \ref{fig:dcf}. 

\begin{figure*}
    \centering
    \includegraphics[scale=0.5]{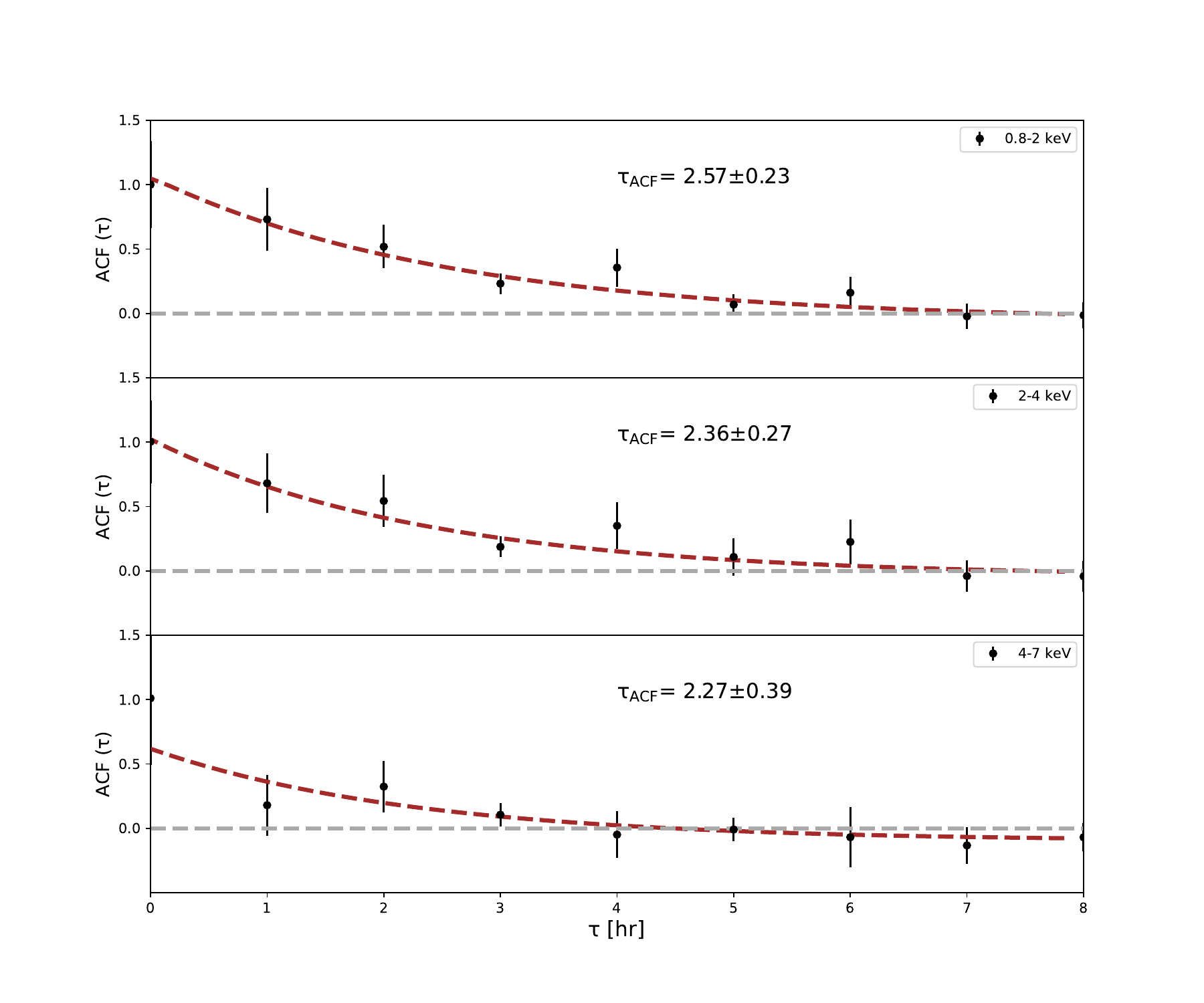}
    \includegraphics[scale=0.7]{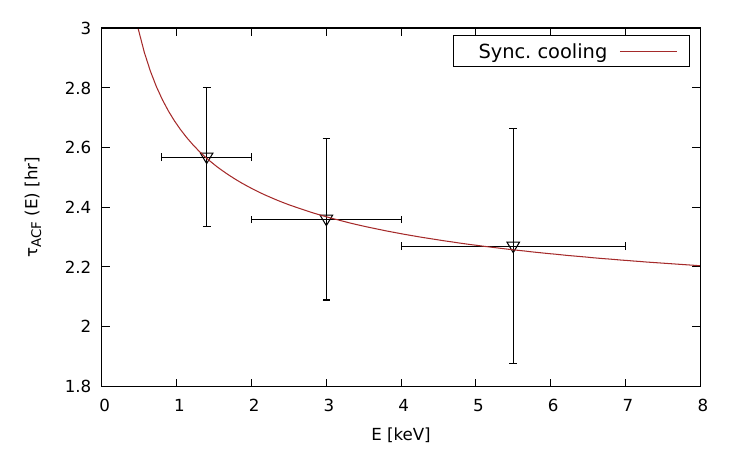}

    \caption{Top plot: Discrete autocorrelation functions (ACFs) of the X-ray flux in three energy channels: L = (0.8 - 2) keV, M = (2 - 4) keV, H = (4 - 7) keV. The ACFs have been fitted with a symmetric constant + exponential. Bottom plot: The widths of the ACFs of the light curves in the three energy ranges, as a function of synchrotron photon energy. The solid curve is the best fit of this energy dependence assuming it is caused by synchrotron + external-Compton cooling.}
    \label{fig:acf}
\end{figure*}

\subsection{Source parameter estimates:}

\citealt{Bottcher_2003} introduced model-independent approaches for estimating the co-moving magnetic field within the source, based on observational results from the BL Lacertae multiwavelength campaign in 2000. Here, we use the SXT light curve to constrain  
the magnetic field within the jet, employing their method. In this procedure, we determine the average rise and decay time scales of variability by estimating the width of the autocorrelation functions (ACFs) of the light curves. Initially, we divide the light curve into three distinct energy channels: 0.8-2 keV, 2-4 keV, and 4-7 keV. Subsequently, we compute the discrete ACFs for these three light curves and fit them using an exponential function of the form expressed as
\begin{equation}
\centering
 ACF({\tau}) = F_{0} + F_{1} exp\bigg(\frac{\tau - \tau_{peak}}{\tau_{ACF}}\bigg)
\end{equation}
where, $\tau_{ACF}$ denotes the decay time, F$_{1}$ indicates the peak value of the ACF observed at time $\tau_{peak}$, and F$_{0}$ is a constant offset. The results are plotted in the top panel of Figure \ref{fig:acf}. 
The `$\tau_{ACF}$' displayed in each panel represents the best-fitted $\tau_{ACF}$ value and is plotted against the photon energy, as shown in the bottom panel of Figure \ref{fig:acf}. The plot suggests a decreasing trend of the variability time scale with increasing photon energy. Clearly, the statistical uncertainties associated with these measurements are substantial, making it challenging to robustly constrain the functional dependence of the ACF widths on photon energy. However, we propose a method to use the decreasing ACF width with increasing photon energy for an independent magnetic field constraint.  

Assuming that the rise time of short-term flaring is independent of energy (i.e., it is primarily determined by light-crossing time constraints rather than an energy-dependent acceleration time scale), 
the width of the ACF can not be smaller than the cooling time scale $\tau_{cool} (E)$ of electrons responsible for emission at photon energy $E$, i.e., $\tau_{\rm ACF}  (E) \gtrsim \tau_{\rm cool} (E)$ in this scenario.  
If electron cooling is primarily due to synchrotron and/or external Compton cooling processes, then the synchrotron+EC cooling time (in the observer’s frame) can be expressed as 

\begin{equation}
\tau_{\rm cool, sy} (E) = 2.9 \times 10^3 \, \delta^{-1/2} 
B_{\rm G}^{-3/2} \, (1 + k)^{-1} \, E_{\rm keV}^{-1/2} 
\; {\rm s}.
\label{tau_sy}
\end{equation}
where B denotes the magnetic field strength in the emission region measured in Gauss, $k\equiv u_{\text{ph}} /u_{B}$ represents the ratio of the energy density in the target photon field for Compton scattering to the magnetic field energy density (equal to the Compton dominance, $C \equiv \nu F_{\nu}^{C}/\nu F_{\nu}^{\rm sy}$ if Compton scattering occurs in the Thomson regime), and $E_{\rm keV}$ stands for the characteristic photon energy emitted by radiative electrons, measured in keV. By fitting a function $\tau_{\rm ACF} (E) = \tau_0 + \tau_1 \, E_{\rm keV}^{-1/2}$ to the energy dependence of the decay timescale as depicted in bottom panel of Figure \ref{fig:acf} (solid line), we obtain a best-fit value of $\tau_1 = (2628 \pm 165)$~s, implying $B \gtrsim 0.27 \, (\delta/15)^{-2/3}$ Gauss for $k\,\sim$ 1 (as $C \sim 1$ typically for BL Lac). This magnetic field constraint is consistent with the values required for the SED and lightcurve modeling study (see \S 4). 

\subsection{Spectral variations:}
\begin{figure}
    \centering
    \includegraphics[scale=0.5]{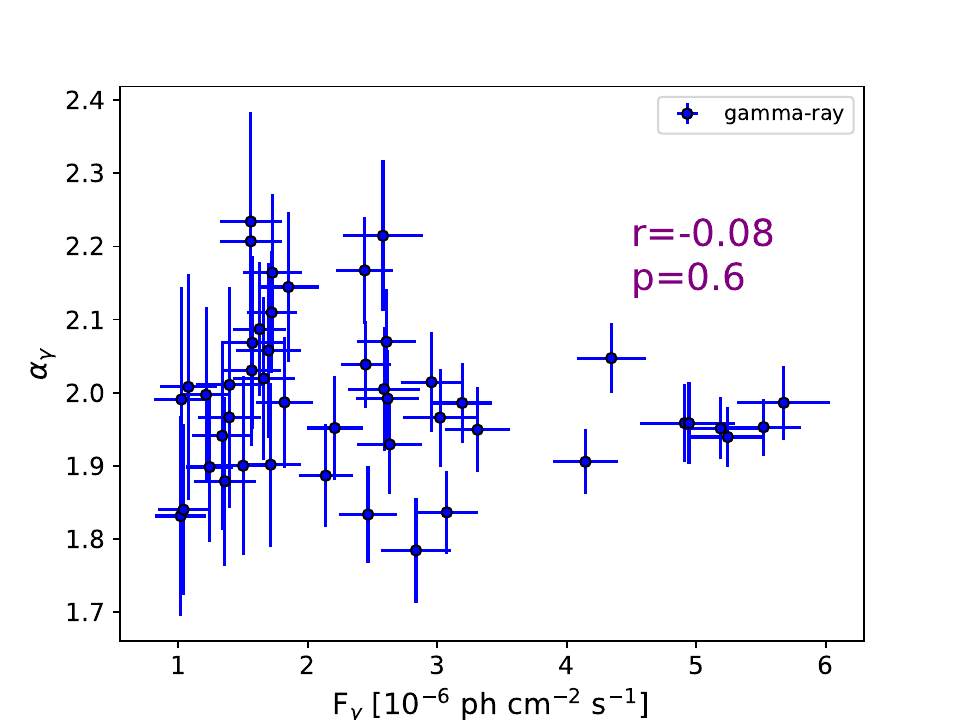}
    \includegraphics[scale=0.5]{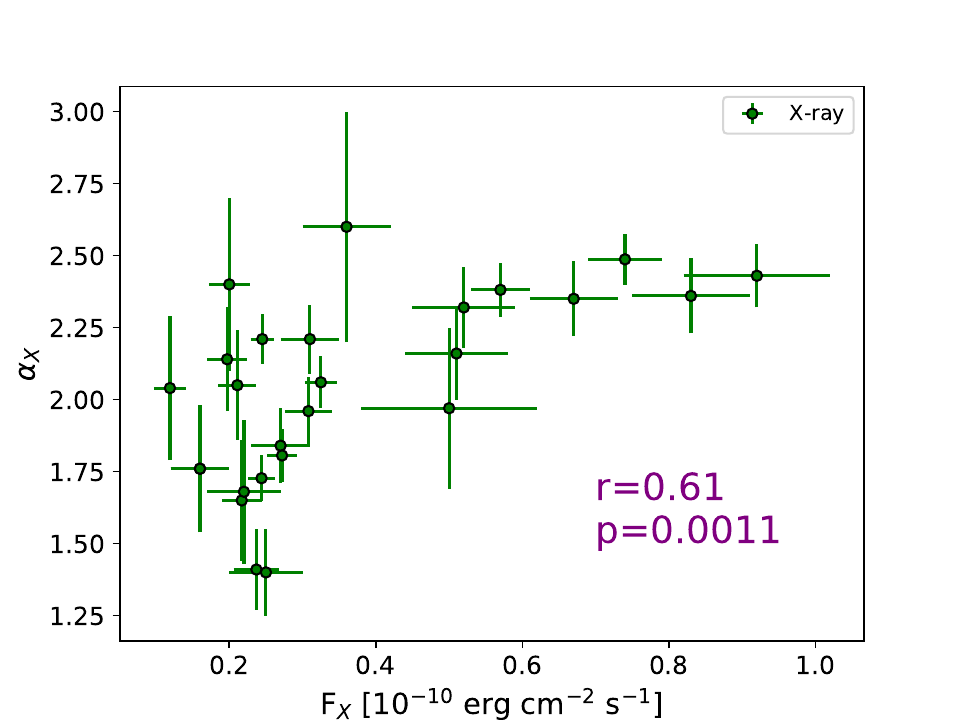}
   
    \caption{Top: scatter plot depicting the 1-day binned $\gamma$-ray flux (100 MeV - 300 GeV) against the spectral index ($\alpha$). Bottom: scatter plot between X-ray flux (0.3-10 keV) and spectral index ($\alpha$).}
    \label{fig:spear-corr}
\end{figure}

To investigate the spectral variability of the source, we performed a Spearman's rank correlation analysis to examine the relationship between the parameters defining the spectral shape and the integrated flux using the gamma-ray and X-ray light curves shown in Figure \ref{fig:mwl_lc}. Figure \ref{fig:spear-corr} presents a scatter plot showing the relationship between the one-day binned $\gamma$-ray flux (F$_{\gamma}$) and the best-fit spectral index $\alpha_{\gamma}$, as depicted in the left plot of the figure. The Spearman's rank correlation analysis between F$_{\gamma}$ and $\alpha_{\gamma}$ resulted in a rank coefficient of r = -0.08, with a null-hypothesis probability of P = 0.6. The high value of P suggests that the correlation between F$_{\gamma}$ and $\alpha_{\gamma}$ is inconclusive.

On the other hand, the Spearman's rank correlation analysis between X-ray flux, F$_{X}$, and the spectral index $\alpha_{X}$,  yielded a correlation coefficient of r = 0.62 with a p-value of 4.2$\times$10$^{-4}$, suggesting a strong correlation between F$_{X}$ and $\alpha_{X}$, as depicted in the plot on the right side of Figure \ref{fig:spear-corr}. This suggests that during the X-ray flare, the photon index softens, indicating a softer-when-brighter trend. A similar behavior of the X-ray flux of BL Lac has been observed in previous studies 
(\citealt{Wehrle_2016, Weaver_2020, Prince_bllac_2021, Dammando_2022}) and attributed to the synchrotron component becoming dominant at soft X-rays during X-ray bright states. This will be further discussed in section \ref{summary}. 

\section{Modeling of SEDs and lightcurves :}

In this section, we present the modeling of the broad-band emission of BL Lacertae. We utilize a time-dependent leptonic diffusive shock acceleration and radiation transfer model. The model description including the numerical techniques employed to solve the time-dependent electron continuity equation and photon transport equations, can be found in \cite{BB_2019}. Here, we will briefly outline the key features of this model.

The model assumes a scenario where mildly relativistic shocks with oblique magnetic-field configurations propagate through the jet of a blazar. It consists of two zones: a small acceleration zone, where diffusive shock acceleration is active, and a larger radiation zone, where shock-accelerated electrons are injected in a time-dependent manner.
These shocks naturally occur when two ultrarelativistic MHD flows collide. In the conventional shock acceleration scenario, the first-order Fermi acceleration process is facilitated by stochastic pitch-angle diffusion of electrons spiraling along magnetic field lines. A convenient parameterization of the mean free path (mfp) of electrons for pitch-angle scattering (PAS) is given by, ${\lambda}_{\rm pas} = \eta_0 \, r_g \, p^{\alpha - 1}$, where $\eta_0$ and $\alpha$ are derived from the Monte Carlo simulations by \cite{SB_2012}, which are briefly discussed in the subsequent paragraph, and 
$r_g$ = ${pc}/{qB}$ denotes the gyroradius of an electron, with p and q representing its momentum and charge, respectively. As $r_g$ scales linearly with $p$, the mean free path scales as $\lambda_{\rm pas} \propto p^{\alpha}$. 

This model incorporates hybrid thermal and nonthermal electron spectra generated through Monte Carlo simulations (\citealt{SB_2012}) of diffusive shock acceleration by mildly relativistic, oblique shocks. The generated electron spectra serve as an injection term into simulations of subsequent radiative cooling of the electrons. Cooling occurs through synchrotron radiation, synchrotron self-Compton (SSC) radiation, and inverse Compton scattering of external radiation fields. 

We estimated the sizes of the Broad Line Region (BLR) and the dusty torus, which can be scaled as $R_{BLR} = 10^{17} {L_{d,45}}^{1/2}$ cm and $R_{IR} = 2.5 \times 10^{18} {L_{d,45}}^{1/2}$ cm \citep{Ghisellini_2011},
where $L_{d,45}$ = 0.6 is the disk luminosity in units of $10^{45}$ erg s$^{-1}$. The calculated values are $R_{BLR} = 7.7 \times 10^{16}$ cm and $R_{IR} = 1.8 \times 10^{18}$ cm. These results indicate that the emission region at a distance of approximately $3.0 \times 10^{17}$ cm lies outside the BLR but within the dusty torus. Consequently, for the inverse Compton process, we consider the IR emission from the dusty torus as the dominant target radiation field for external Compton scattering. We assume that the emission from the dusty torus follows a black body distribution with a temperature of $\approx$ 1000 K, and the energy density of the IR photon field is estimated to be around $5 \times 10^{-5}$ erg/cm$^3$ in the AGN rest frame. 
The resulting time-dependent radiative output due to the hybrid electron distributions, is evaluated using radiation transfer schemes described in detail in \cite{bottcher_chiang_2002}, \cite{B13}, and \citealt{BB_2019}.

The emission components were calculated assuming isotropic distributions of the relativistic electrons within the jet frame. They were subsequently  Doppler-boosted to the observer's frame using the bulk Lorentz factor ($\Gamma$) of the jet and the observer's viewing angle ($\theta_{\rm obs}$) relative to its axis. In the modelling, specific values are adopted for certain parameters: a shock speed of $v_s = 0.71$~c, a magnetic field obliquity to the shock normal of $\Theta_{\rm Bf1} = 32^o.3$, an upstream plasma temperature of $5.45 \times 10^7$~K, and a velocity compression ratio of $r = 3.71$ (\citealt{BB_2019}). These choices effectively represent the environment of a strong, subluminal, mildly relativistic shock. The SED and lightcurves are primarily governed by the shock parameters $\eta_0$ and $\alpha$, the power dissipated by the shock and transferred to relativistic electrons (referred to as ``injection luminosity", $L_{\rm inj}$ hereafter), the strength of the magnetic field (B),  
the bulk Lorentz factor $\Gamma$, and the radius $R$ of the emission region. During modeling of various flux states, all these parameters remain free except for $\Gamma$ and $\theta_{\rm obs}$, which were fixed at values of 15, and 3.82$^o$, respectively, the latter representing the critical (superluminal) angle, $\theta \approx 1/\Gamma$, for which $\delta = \Gamma$. The details of the other parameters are discussed in the following subsections.

\subsection{SED modeling, results and interpretation:}
To conduct broadband SED modeling of BL Lac across various flux states, we chose three distinct time intervals: one representing a quiescent state (Q state) spanning MJD 59441 to 59443, and two corresponding to flaring states covering MJD 59406 to 59407 (F1 state) and MJD 59430 to 59431 (F2 state), as illustrated in Figure \ref{fig:mwl_lc}. These selections were made to ensure that each segment encompasses simultaneous observations across gamma-ray, X-ray, UV, and optical wavelengths. 

Starting with the parameters derived in previous sections and choices of $R \sim 10^{16}$~cm, and $B \sim 0.3$~Gauss, we have done a series of simulations with the one-zone time-dependent shock-in-jet leptonic code elaborated in the previous section.
We initiate our fitting procedure by configuring a quiescent state, which accurately reproduces the low-state SED of BL Lacertae, by letting the code adjust self-consistently to an equilibrium state with constant shock-acceleration parameters. Subsequently, utilizing the parameters from our quiescent state fitting, we proceed to model SEDs for various flaring scenarios by changing the shock-acceleration parameters in a time-dependent manner and extracting snap-shot SEDs during the action of these modified parameters. 

The solid curves depicted in Figure \ref{fig:sed} represent the best-fit model SEDs corresponding to the low-state, F1, and F2 states, respectively. The relevant fit parameters can be found in Table \ref{sed_parameters}.
The dominant change of parameters between the quiescent and
the flaring state is given by a hardening of the electron
spectrum through a significant change of the PAS mean-free-path momentum-scaling index $\alpha$. In addition, a significant change of the injection luminosity L$_{jet}$ and slight changes in the magnetic field are required.

\subsubsection{Recurrence of the HBL component}
Flux variations across various bands, including optical/UV, X-rays, and $\gamma$-rays, have been consistently observed in BL Lacertae over an extended period. \citealt{Sahakyan_2022} conducted a comprehensive analysis of the broadband emission from BL Lac over nearly 13 years, from August 2008 to March 2021. They conducted the modelling by fitting leptonic models incorporating synchrotron self-Compton and external Compton components to 511 high-quality and quasi-simultaneous broad-band SEDs. This study
illustrates that the source exhibits highly variable emission across all frequencies, often accompanied by spectral changes. Notably, during the brightest flare on MJD 59128.18 (October 6, 2020), significant X-ray spectral variability was observed. This was attributed to a notable shift in the synchrotron peak location from its typical location at $\nu_{\rm sy} \sim 10^{14}$~Hz to $\sim 10^{16}$~Hz, well into the regime typically associated with HBLs. \citealt{Prince_bllac_2021} and \citealt{DAmmando_2022_bllac} investigated this phenomenon, marking it as an exceptional occurrence observed for the first time during this particular flare. We obtained similar results when studying the source from July 2021 to August 2021, including the brightest flare observed in 2021. In Figure \ref{fig:sed}, the peak of the synchrotron emission during the low flux-state is recorded at $2 \times10^{14}$~Hz, while during flare F1 (MJD 59406-59407), it reaches approximately $2 \times10^{15}$~Hz. This indicates the persistence of this extraordinary behavior, even during the 2021 flare. However, such behavior is evidently not replicated by flare F2, where the synchrotron peak is around $3 \times10^{14}$~Hz.

\subsubsection{Jet frame luminosities:}
Modeling enables us to evaluate the power (or luminosity) of the jet across different time intervals. 
We calculated the power carried by the electrons, the protons (assuming the presence of one cold proton per electron), and the magnetic field within the jet as
\begin{equation}
L_{jet} = \pi\, R^2 \,{\Gamma^2}\,c \,(U_{e}+U_{p}+U_{B})
\label{jet_lum}
\end{equation}
where U$_{e}$, U$_{p}$, and U$_{B}$ represent the co-moving electron energy density, proton energy density, and the magnetic field energy density, respectively. The calculated power is presented in Table \ref{sed_parameters}.   
The total jet power is in the ranges $(3.38 - 11.2) \times 10^{45}$~erg/s, which is below the Eddington luminosity of $4.75 \times 10^{46}$~erg/s for a supermassive black hole mass of $3.8 \times 10^{8} \, M_{\odot}$ in BL Lac (\citealt{Wu_2009, Titarchuk_2017}). Our fit results require the power in the injected electrons dominates over that in the magnetic field, indicating a low magnetization of the jet, which favours the efficiency of the formation of strong shocks. 

\subsubsection{Lightcurve modeling:}
\begin{figure}
    \includegraphics[scale=0.30
    ]{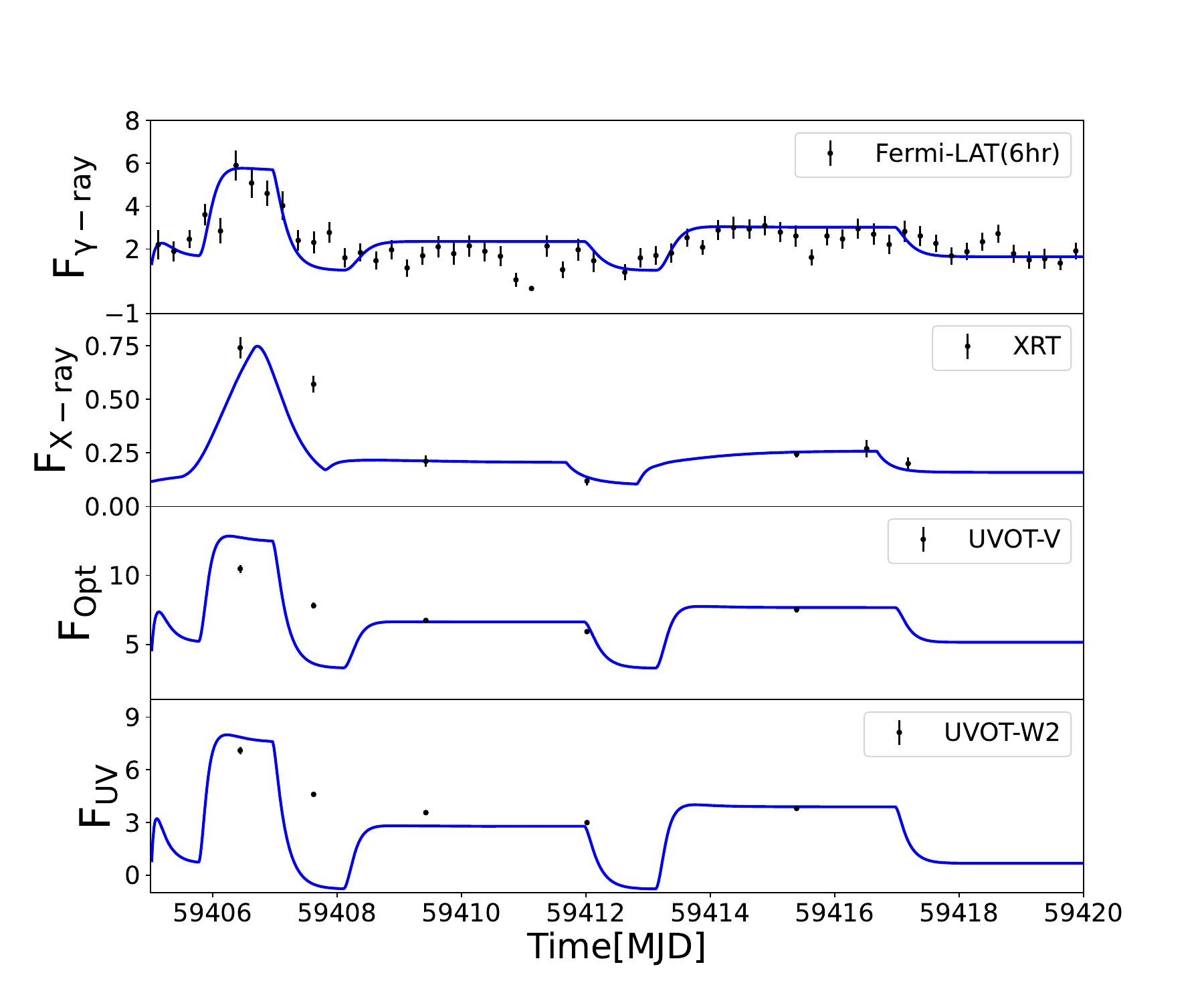}

    \caption{LC Modelling results with three shocks. }
    \label{fig:lcmodeling}
\end{figure}

The multiband light curves shown in Figure \ref{fig:mwl_lc} display distinct individual flares. 
Attempting to model the entire $\sim 2$ months light curve would require the introduction of an unreasonably large number of parameter variations throughout the various variablity features. Therefore, as an illutrative example of the ability of the adopted shock-in-jet model to reproduce both snap-shot SEDs and light curve segments simultaneously, we confine the modeling of light curves to the period from MJD 59400 to MJD 59420, as shown in Figure \ref{fig:lcmodeling}. The flux variations observed across multiple bands indicate that they cannot be solely attributed to a single impulsive particle acceleration event. Instead,
they imply a series of successive shocks occurring at different times throughout the emission region. Our analysis reveals that a sequence of three shocks of varying strength is required to achieve a satisfactory representation of the light curve segments.

\begin{figure*}
    \centering
    \includegraphics[scale=0.5]{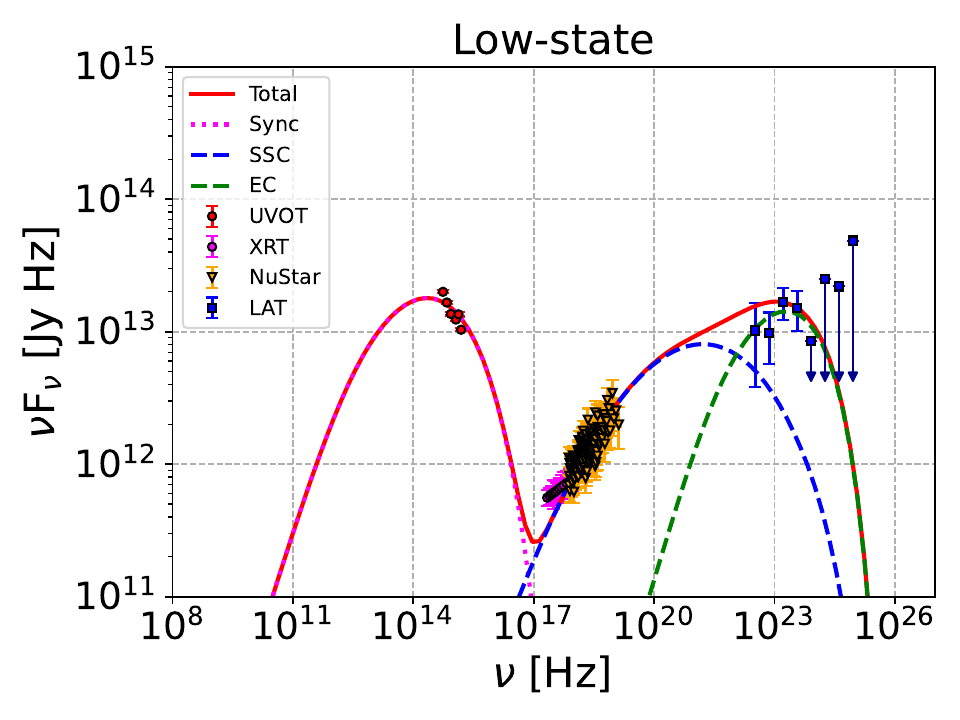}
    \includegraphics[scale=0.5]{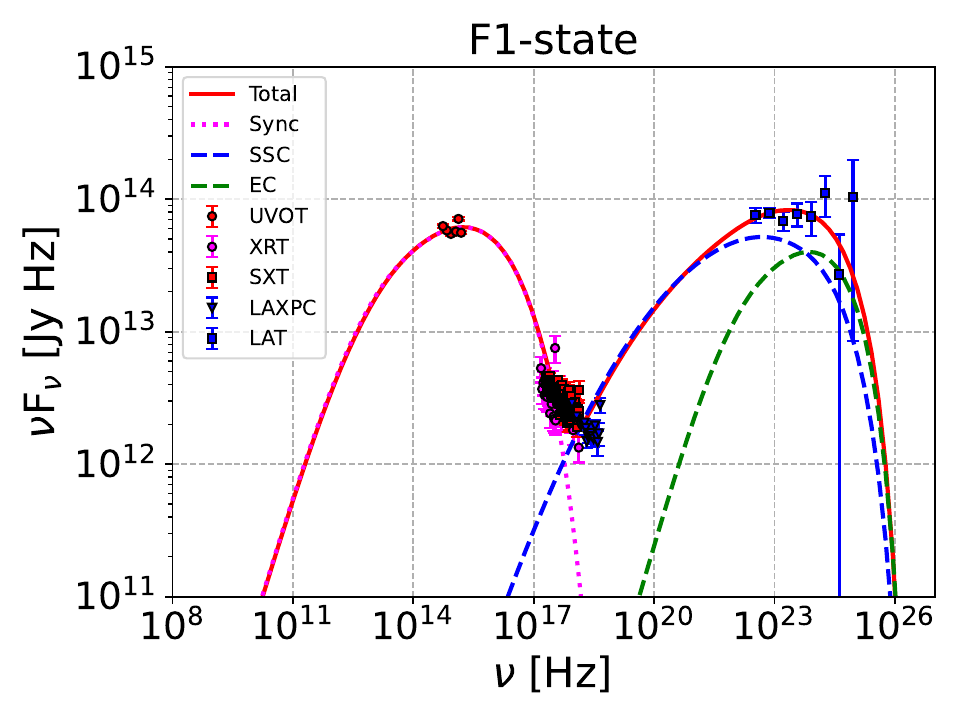}
    \includegraphics[scale=0.5]{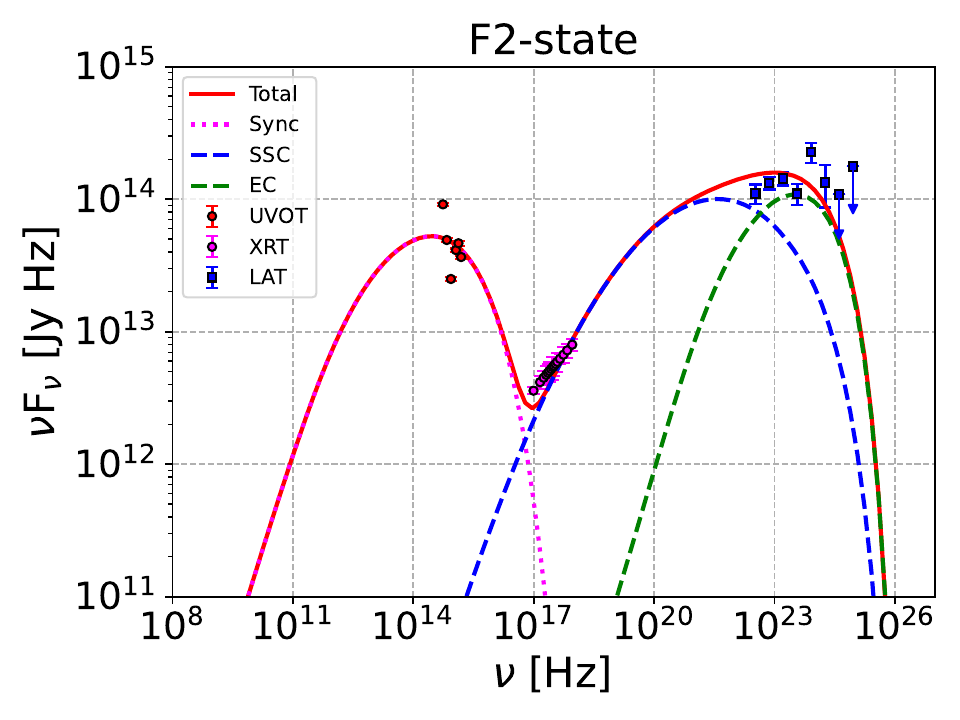}
   
    \caption{Modelling the observed broadband SEDs during the low, F1, and F2 states using the time-dependent shock-in-jet model of \cite{BB_2019}}.
    \label{fig:sed}
\end{figure*}

\begin{figure}
    \includegraphics[scale=0.50,angle=-90
    ]{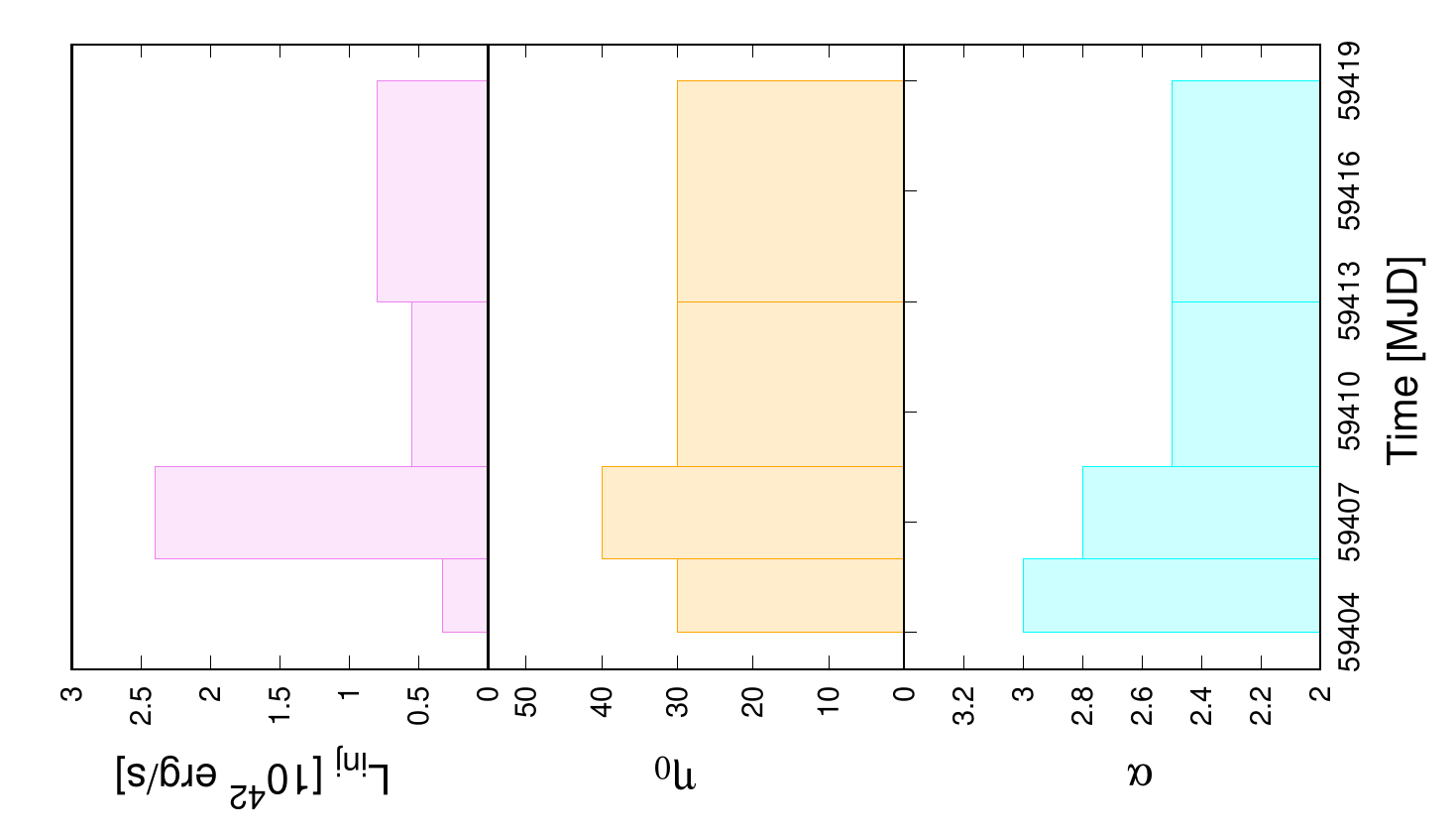}

    \caption{The evolution of the shock parameters obtained from the LC modeling (see Table \ref{3_shocks}, Figure \ref{fig:lcmodeling}) }
    \label{fig:lc_model_params}
\end{figure}

The light curve modeling process begins with the quiescent state configuration (see Table \ref{sed_parameters}, Figure \ref{fig:sed}), which is selected as the reference low state of the initial outburst, and to fit the subsequent outbursts, changes in the properties of the shock parameters are applied. Each of the three shocks traversing the jet is distinguished by an increased injection luminosity and a decrease in the mean free path for pitch angle scattering, indicated by a slightly decreased value of $\alpha$. This adjustment leads to more efficient particle acceleration, contributing to the observed increase in variability amplitude within the light curve. The parameters used for the three shocks in our simulation are listed in Table \ref{3_shocks}. The evolution of the $L_{inj}$, $\eta_{0}$, and $\alpha$ parameters is shown in Figure \ref{fig:lc_model_params}.

We showed that the physical model, based on diffusive shock acceleration by mildly relativistic, oblique shocks within the jet of BL Lacertae, can successfully reproduce the multiband light curves, spanning optical, UV, X-ray, and gamma-ray bands. This marks the first instance where the model accurately captures the multiband temporal variability, including hour-scale variability observed during the flare. Previously, such a model effectively reproduced the X-ray variability of 1ES~1959+650 (\citealt{chandra_2021}) and the minute-scale TeV variability of PKS 2155-304 (\citealt{Tan_2024}).

\section{\label{summary}SUMMARY AND DISCUSSION}

In early July 2021, the blazar BL Lacertae underwent a multiwavelength outburst, prompting our in-depth investigation into its spectral and light curve characteristics. We utilized data from Swift-XRT/UVOT, AstroSat, NuSTAR, and Fermi-LAT, spanning from July 2021 (MJD 59400) to August 2021 (MJD 59450). The multi-wavelength variability study between the fluxes in different wavebands suggests the fractional variability amplitude (F$_{var}$) is larger in the
$\gamma$-ray  ($\sim 62$~\%) and X-ray ($\sim 60$~\%) bands compared to UV and optical wavelengths.

The source exhibited intense flaring activity, characterized by two significant outbursts denoted as F1 and F2 (Figure \ref{fig:mwl_lc}), occurring during MJD 59406-59407 and MJD 9429.5-59430.5, respectively. The fastest variability halving time-scale in the gamma-ray band (t$_{var}$) was determined to be $\sim$ 5.6 hours. \cite{pandey_2022} reported the first detection of minute-timescale GeV $\gamma$-ray variability in BL-Lac type blazars, with a flux halving time of approximately 1 minute on April 27, 2021. Additionally, we calculated t$_{var}$ using X-ray observations from the AstroSat-SXT/LAXPC telescope, yielding values of around 1 hour and 1.4 hours, respectively. 
Furthermore, the DCF correlation study across all wavebands indicates that the broadband radiation may originate from a single emission region. Using the gamma-ray variability timescale of approximately 5.6 hour and a Doppler factor ($\delta$) of around 15, we constrained the size of the emission region to be $\lesssim 1.0 \times10^{16}$~cm. The SED model best-fit value of radius is consistent with the model-independent estimation of $R$. 

In this work, we employ a model-independent method to constrain the magnetic field using the width of the autocorrelation function (ACF) of the SXT light curves (0.8-2 keV, 2-4 keV, and 4-7 keV) at different photon energies, assuming that the ACF widths serve as upper limits on the energy-dependent radiative cooling time scales. With this approach, the derived magnetic field strength measures $\gtrsim$ 0.3 Gauss. From the SED modeling, we found B to be around 0.5-0.75 Gauss, which is consistent with the derived lower limit, but slightly lower than the magnetic field values  
reported by other authors based on earlier multiband campaigns on this source (\citealt{BR_2004, Prince_bllac_2021, Sahakyan_2022}).
Based on the magnetic field estimate, and the approximate location of the synchrotron peak of the SED of BL Lacertae, we were able to estimate the peak energy of the electron energy distribution in the synchrotron-emitting region.
The characteristic frequency of the underlying electron distribution responsible for the synchrotron peak frequency in the SED is given by
$\nu_{\rm sy} (E) = 4.2 \times 10^6 \, {\gamma_{\rm p}}^{2} 
B_{\rm G} \, {\delta} \, /\rm(1+z)
\; {\rm Hz}.$
Using the parameter values such as $\delta$ = 15, redshift of the source of z = 0.069, the lower limit on the magnetic field strength of $B_{\rm G}\, \gtrsim \, 0.3$ Gauss, 
and approximate location of the synchrotron peak frequency ($\nu_{\rm sy}$) for the F1 state at around $\sim 2 \times \rm 10^{15}$~Hz, an upper-limit of the Lorentz factor at the peak of the electron distribution was estimated as $\gamma_{\rm pk} \rm \lesssim 1.1\rm\times\rm 10^{5}$. 

We examined the spectral variability of the source using a Spearman’s rank correlation analysis on gamma-ray and X-ray light curves (see Figure \ref{fig:mwl_lc}). 
A correlation between gamma-ray flux and its corresponding photon spectral index is not clearly established. However, the analysis of the X-ray flux and its corresponding spectral index showed a softer-when-brighter trend, consistent with previous studies. This softening in the X-ray spectrum appears to be related to a change of the peak frequency of the synchrotron component.

Comparing these results with SED modeling, we found that when the X-ray flux ($F_X$) is high ($\sim 7.4 \times 10 ^{-11}$ erg cm$^{-2}$ s$^{-1}$) and associated with a soft component ($\alpha_{x}$ $\sim$ 2.5), the peak frequency of the low-energy component of the SED ($\nu_{syn}$) reaches  frequencies of 2$\times$10$^{15}$ Hz, placing BL Lac temporarily in the domain of HBL blazars. As the X-ray flux decreases to $\sim$ 1.7$\times$10$^{-11}$ erg cm$^{-2}$ s$^{-1}$, with $\alpha_{x}$ $\sim$ 1.77, the peak frequency $\nu_{syn}$ is recorded as 2$\times$10$^{14}$ Hz, placing BL Lac into the IBL category.
However, during flare F2, when $F_x$ increases to $\sim$ 9.25$\times$10$^{-11}$ erg cm$^{-2}$ s$^{-1}$, with $\alpha_{x}$ $\sim$ 2.4, the peak frequency $\nu_{syn}$ is recorded as $3 \times 10^{14}$~Hz, indicating that the source remains in the IBL category. 
The presence of the HBL component during flares has also been previously noted in this source by \citealt{DAmmando_2022_bllac} and \citealt{Prince_bllac_2021}. 

\begin{figure}
    \includegraphics[scale=0.5]{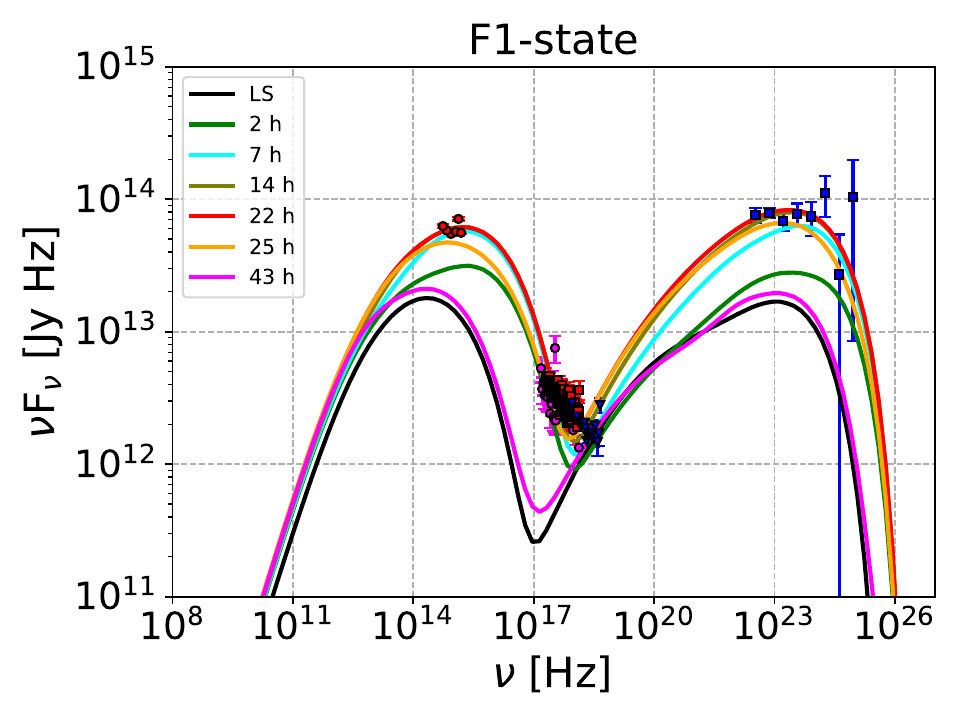}
    \caption{Snapshots of the fitted SEDs for the F1 state at various times during the simulation.}
    \label{fig:snap_sed}
\end{figure}

The key novelty of this work lies in demonstrating the capability of the time-dependent leptonic model, based on diffusive shock acceleration by mildly relativistic shocks within the source's jet, to explain both the spectral and temporal behavior of the source. This provides valuable insights into the particle acceleration mechanisms responsible for the variations in the emitted radiation.  The broadband SED of BL Lacertae during its quiescent state (QS), as well as during flaring periods F1 and F2, along with a section of the multiband light curves spanning from MJD 59400 to MJD 59420, have been effectively fitted using this model. The interaction between shock acceleration, self-consistent radiative processes (synchrotron, synchrotron self-Compton, and external Compton) and cooling of particles within the emission region produced flux and spectral variability patterns consistent with observations. With this model, the variability patterns of the source during flares can be plausibly reproduced predominantly by increasing the number of radiating non-thermal electrons generated by the shocks, thereby enhancing the synchrotron, SSC and EC emission at almost same rate. The model can account for the hour-scale variability observed during the flare with a moderate and typical value of $\delta$ = 15, and it explains the simultaneous flux lightcurves from the optical to the high-energy gamma-ray bands including the F1 flare. Snapshots of the SEDs during the F1 state are included in Figure \ref{fig:snap_sed}.

We found that multi-wavelength flares may arise due to turbulence generated by shocks, leading to a reduction in the mean free path of electrons for pitch-angle scattering ($\lambda_{pas}$), as indicated by a slight decrease in the value of $\alpha$, thereby enhancing particle acceleration efficiency. Additionally, an increase in particle injection luminosity, along with slight variations in magnetic field strength, contributes to flux enhancements. Future high-quality, dense, simultaneous multiband light curves will help constrain the theoretical model more precisely.

\begin{table}[t]
    \centering
     \caption{{ \footnotesize \label{tab:obslog} Details of the data used for the present study.}}
    {\footnotesize \begin{tabular}{clll}
    \hline
         S.N. & Instrument & Total Exposure & Epoch of Observations \\
         \hline
         1 & {\it Fermi}-LAT & -  &  59400.0 - 59450.0  \\
         2 & {\it Swift} & 411.2 ks &  59400.0 - 59450.0 \\
         3 & {\it AstroSat} & 20.0 ks &  59406.0 - 59407.0 \\
         4 & {\it NuStar} & 20.0 ks &  59442.5 - 59443.5 \\

         \hline  
    \end{tabular}}
  \end{table}
\begin{table}
	\caption{Fractional variability ({\it $F_{var}$}) in different wavebands.}
	\begin{tabular}{l c }
		\hline
		Waveband     			& {\it $F_{var}$} (MJD 59400 - 59450)\\
		\hline
		
{\it Fermi}-LAT (0.1-300GeV) &  $ 0.62 \pm  0.01 $\\
\emph{Swift}-XRT (0.3-10keV) & $ 0.58 \pm 0.03 $\\
UVOT band-B  & $ 0.34 \pm 0.006 $\\
UVOT band-U  & $ 0.39 \pm 0.006 $\\
UVOT band-V  & $ 0.40 \pm 0.007 $ \\
UVOT band-W1  & $ 0.40 \pm 0.006 $\\
UVOT band-W2  & $ 0.41 \pm 0.006 $\\
UVOT band-M2 & $ 0.51 \pm 0.009 $\\

	\hline 
	\end{tabular}
	\label{tab:fvar}
\end{table}

\begin{table*}[t]
  \centering
  \caption{{ \footnotesize 
  \label{sed_parameters} Model fit parameters obtained by fitting the broadband spectrum of low-state, F1-state and F2-state of BL-Lac with the time-dependent shock-in-jet model (\citealt{BB_2019}))}}
{\footnotesize \begin{tabular}{cllll}
    \hline
    
 &&Jet-frame parameters&&\cr
    \hline
Name & Symbol/units & LS & F1 & F2 \cr
\hline
Injection mean free path & $\eta_0$ & 30 & 40 & 30\cr
Diffusion index & $\alpha$ & 3.0 & 2.5 & 2.8\cr
Electron injection luminosity & $L_{\rm inj}$ [erg/s] & $3.3 \times 10^{41}$ & $1.2 \times 10^{42}$ & $2.2 \times 10^{42}$\cr
Magnetic field &$B$ [G] & 0.75 & 0.75 & 0.5\cr
Emission region size & $R$ [cm] & $1.0 \times 10^{16}$  & $1.0 \times 10^{16}$ & $1.0 \times 10^{16}$\cr
\hline
&Derived parameters &  &  &  \cr
\hline
Jet power in electrons &$L_{e}$ [erg/s]  & 
$2.06 \times 10^{44}$ & $1.56 \times 10^{44}$ & $2.24 \times 10^{44}$\cr

Jet power in protons & $L_{p}$ [erg/s]  & $3.18 \times 10^{45}$ & $3.18 \times 10^{45}$ & $1.17 \times 10^{46}$\cr

Jet power in magnetic field &$L_{B}$ [erg/s]  & $4.75 \times 10^{43}$ & $4.75 \times 10^{43}$ & $2.11 \times 10^{43}$\cr

Total jet power & $L_{jet}$ [erg/s]  & $3.43 \times 10^{45}$ & $3.38 \times 10^{45}$ & $1.12 \times 10^{46}$\cr

\hline
  \end{tabular}}
  \end{table*}

\begin{table}
\caption{\label{3_shocks}Parameter variations for the fits to MWL light curves.}
\begin{tabular}{ccccc}
\hline
Parameter [units] & $L_{\rm inj}$ [erg/s] & $\eta_0$ & $\alpha$ & $B$ [G]\cr
\hline
Quiescence & $3.3 \times 10^{41}$ & 30 & 3.0 & 0.75 \cr
Shock 1 & $2.4 \times 10^{42} $   & 40 & 2.8 & 0.75\cr
Shock 2 & $5.5 \times 10^{41}$    & 30 & 2.5 & 0.75\cr
Shock 3 & $8.0 \times 10^{41}$    & 30 & 2.5 & 0.75\cr

\hline
\end{tabular}
\end{table}

\begin{acknowledgments}
{\bf Acknowledgements:}
We thank the anonymous referee for suggestions helping to improve the presentation. R. K. thanks Dr. Sunil Chandra, for the help in SXT SED analysis. R. K. and M. B. acknowledge support from the South African Department of Science and Innovation and the National Research Foundation through the South African Gamma-Ray Astronomy Programme (SA-GAMMA). 
\end{acknowledgments}

\vspace{5mm}
\facilities{Fermi(LAT), Swift(XRT and UVOT), AstroSat (SXT/LAXPC), NuStar}

\software{Fermitools (\url{https://fermi.gsfc.nasa.gov/ssc/data/analysis/scitools/})\\
\texttt{HEAsoft}-v 6.27\footnote{https://heasarc.gsfc.nasa.gov/lheasoft/download.html}\citep{2014ascl.soft08004N},
\texttt{XSPEC}\footnote{https://heasarc.gsfc.nasa.gov/xanadu/xspec/}\citep{Arnaud_1996},%
\texttt{FTOOLS}\footnote{https://heasarc.gsfc.nasa.gov/ftools/xselect/}\citep{1995ASPC...77..367B}}



\bibliography{main}{}
\bibliographystyle{aasjournal}

\end{document}